\documentclass[useAMS,usenatbib,usegraphicx]{mn2e}
\usepackage{times}
\usepackage{amsfonts,amssymb}
\usepackage{color}

\title[Covering Factor of Warm Dust in Quasars]
  {The Covering Factor of Warm Dust in Quasars: View from WISE All-Sky
   Data Release}
\author[Ma \& Wang]
  {Xiang-Cheng Ma,$^{1,2}$\thanks{xchma@mail.ustc.edu.cn}
  Ting-Gui Wang$^{1,2}$ \\
  $^1$Department of Astronomy, School of Physical Science, 
	The University of Science and Technology of China (USTC), Hefei, Anhui 230026, P. R. China \\
  $^2$Key Laboratory for Research in Galaxies and Cosmology, Chinese Academy of Science}

\pagerange{\pageref{firstpage}--\pageref{lastpage}}

\newcommand{\LNUV}{L_{\rm NUV}}
\newcommand{\LMIR}{L_{\rm MIR}}
\newcommand{\Lbol}{L_{\rm bol}}
\newcommand{\CF}{\rm CF_{WD}}
\newcommand{\MBH}{M_{\rm BH}}
\newcommand{\FNUV}{F_{\rm NUV}}
\newcommand{\FMIR}{F_{\rm MIR}}

\begin{document}

\pagerange{\pageref{firstpage}--\pageref{lastpage}} \pubyear{2013}

\maketitle

\label{firstpage}

\begin{abstract}
By combining the newly infrared photometric data from the All-Sky Data 
Release of the \textit{Wide-field Infrared Survey Explorer} with the 
spectroscopic data from the Seventh Data Release of the Sloan Digital 
Sky Survey, we study the covering factor of warm dust ($\CF$) for a 
large quasar sample, as well as the relations between $\CF$ and other 
physical parameters of quasars. We find a strong correlation between 
the flux ratio of mid-infrared to near-ultraviolet and the slope of 
near-ultraviolet spectra, which is interpreted as the dust extinction effect. 
After correcting for the dust extinction utilizing the above correlation, 
we examine the relations between $\CF$ and AGN properties: bolometric luminosity
($\Lbol$), black hole mass ($\MBH$) and Eddington ratio ($L/L_{\rm Edd}$).
We confirm the anti-correlation between $\CF$ and $\Lbol$. Further we find that 
$\CF$ is anti-correlated with $\MBH$, but is independent of $L/L_{\rm Edd}$.
Radio-loud quasars are found to follow the same correlations as for radio-quiet quasars.
Monte Carlo simulations show that the anisotropy of UV-optical continuum of 
the accretion disc can significantly affect, but is less likely to dominate 
the $\CF$--$\Lbol$ correlation.
\end{abstract}

\begin{keywords}
catalogues--galaxies: active--quasars: general--infrared: galaxies
\end{keywords}

\section{Introduction}

The terminology `active galactic nucleus' (AGN) generally refers to energetic
phenomena in the central regions of galaxies that can be attributed to the
accretion of gas onto supermassive black holes. Due to its angular momentum,
the in-falling gas forms a disc around the black hole and slowly spirals
inwards as friction torque transports the angular momentum out.
During this process, gas in the disc is heated to a few $10^5$~K and produces
strong UV to optical continuum. Strong UV radiation ionizes gas surrounding
the black hole, giving rise to prominent broad emission lines within one
parsec and narrow emission lines further out. These emission components are
further modified by dust in a torus outside the broad line region (BLR) and in 
the interstellar medium of galaxies. Dust absorbs and scatters UV and optical
light and re-emits at infrared (IR) bands, altering spectral energy distribution
(SED) as well as causing anisotropic obscuration. In the past thirty years,
it has been established that a large portion of observed diversities, such
as presence or absence of broad lines and prominent non-stellar continuum,
can be attributed to the anisotropic obscuration of the dusty torus
(see \citealt{antonucci.1993} for a review).

The study of dusty torus is of great significance in our understanding of 
physical processes of AGNs. The torus is a natural reservoir of gas supply 
to the accretion disc, and thus provides a possible connection between
the accretion disc at sub-parsec scale and the more extended nuclear disc
\citep[e.g.][]{krolik.begelman.1986,hopkins.etal.2005,di.matteo.etal.2005}.
It has been argued that the torus is a smooth continuation of the BLR and
the boundary between them is determined only by dust sublimation
\citep[e.g.][]{elitzur.2008}. So the torus may define the extension of BLR
\citep{netzer.laor.1993}. Also, dusty torus may be the source of outflows 
on the scales of parsecs. The anisotropic obscuration of ionizing continuum
causes an ionization cone structure in the narrow line region, which has
been detected in many obscured nearby Seyfert galaxies.

Due to its great importance, the geometry, column density distribution and the 
composition of dusty torus have been subjects of intensive study in the past 
thirty years. The average covering fraction of torus could be constrained from 
the number ratios of obscured to unobscured AGNs in the local universe. 
For nearby Seyfert galaxies and radio galaxies, \citet{lu.etal.2010} reported a 
Type 2 to Type 1 ratio around 2:1 to 3:1. This suggests that an average opening 
angle of the dust torus is about 45 degree, which is basically consistent with 
that derived from quasi-steller radio sources \citep{barthel.1989} and direct 
imaging of ionizing cone in many nearby Seyfert galaxies 
\citep[e.g.][]{wilson.ulvestad.1983,pogge.1988,sako.etal.2000}.
While in the literature, \citet{lawrence.elvis.2010} suggested this ratio to 
be $\sim$1:1, and so did \citet{reyes.etal.2008} by reporting approximately 
equal space densities of obscured and unobscured quasars.
Yet a consensus has still to be reached on whether and how the opening angle
depends on other properties of AGNs. \citet{lawrence.1991} suggested that the
fraction of type I AGNs increases with the increase of the nuclear luminosity.
Similar results were obtained by \citet{simpson.2005} and \citet{hao.strauss.2004} 
for a large sample of AGNs derived from Sloan Digital Sky Survey (SDSS), and by
\citet{hasinger.2008} from deep X-ray surveys. However, \citet{lu.etal.2010}
did not find such a correlation for radio loud AGNs by taking into account of 
various selection effects, and nor did \citet{lawrence.elvis.2010} for radio
quiet AGNs. This inconsistency is largely due to the correction of complicated
selection effects in dealing with the AGNs in each luminosity bin.

The infrared emission is a powerful probe of the dusty torus. With a maximum
temperature around $\sim2000$~K, the emission of dusty torus is mainly at IR 
bands. The total IR luminosity from the torus is a measurement of the amount of
optical to UV light absorbed by the torus, while individual features are used
to probe the composition of grains and their distribution. The \textit{Spitzer}
Space Telescope and more recent \textit{Herschel} provide very rich data-sets,
which allow us to study different features and parameters of the dusty torus.
Meanwhile, large unbiased mid-infrared surveys allow to study the statistical
properties of the dust torus. One of the problems that we are primarily concerned 
is the dust covering factor (CF) and its correlations with AGN properties, such as 
bolometric luminosity, black hole mass and accretion rates. In the literature, 
several authors have suggested that the CF decreases with increasing bolometric
luminosity ($\Lbol$) \citep[see, e.g.][]{maiolino.etal.2007,treister.etal.2008}.
However, these studies used either a relative small sample, or a combination
of a low-redshift, low-luminosity sample with a high-redshift, high-luminosity
sample. In the latter case evolution effect cannot be excluded.

Recently, the \textit{Wide-field Infrared Survey Explorer} 
\citep[WISE;][]{wright.etal.2010} provided a large amount of photometric data 
at near- and mid-infrared bands that can be used to study the IR emission as 
well as dust CF of AGNs for a quite large sample. For example, recent works of 
\citet{mor.trakhtenbrot.2011} and \citet{calderone.etal.2012} reported that 
CF anti-correlates with $\Lbol$ by studying large magnitude-limited samples.
In this paper, we will utilize the most recent WISE All-Sky Data Release,
combined with the spectroscopic data from the Seventh Data Release of the
Sloan Digital Sky Survey \citep[SDSS/DR7;][]{abazajian.etal.2009}, to study 
the warm dust emission of 16275 quasars in a redshift range of $0.76<z<1.17$
(including $\sim10\%$ radio loud quasars). 
The sample is described in \S~\ref{sample}. We fit the SDSS spectra with 
power-law function to derive the flux in the near-ultraviolet (NUV) range 
of rest frame wavelength from 2000~\AA~to 4000~\AA~($\FNUV$). This NUV band 
is close to the peak SED of the big blue bump \citep{zheng.etal.1997,shang.etal.2011}. 
We also calculate the flux in near- to mid-infrared (MIR) range of rest frame 
wavelength from 3~$\mu$m to 10~$\mu$m ($\FMIR$, \S~\ref{flux}), which is dominated 
by warm, and a small portion of hot, dust. We find a strong correlation between 
$\FMIR/\FNUV$ and NUV spectral slope $\alpha$, which can be best explained as dust 
extinction and reddening in NUV (\S~\ref{extinction}). After correcting for the 
dust extinction, we estimate the covering factor of warm dust $\CF$, and examine 
its relations with other AGN properties (\S~\ref{relation}). In the end, we compare 
our results with previous works and investigate how anisotropic continuum can affect 
the observed $\CF$--$\Lbol$ correlation (\S~\ref{discussion}). 

In this paper, we use a flat cosmology with $H_0=70\,\rm km\,s^{-1}\,Mpc^{-1}$, 
$\Omega_{\rm M}=0.3$ and $\Omega_{\rm \Lambda}=0.7$.

\section{The Sample and Measurement of Flux}

\subsection{Sample Description}
\label{sample}

We construct our quasar sample by cross-correlating the SDSS DR7 quasar
catalog \citep{schneider.etal.2010} with the WISE All-Sky Data Release catalog.
This leads to a sample of 101639 quasars with reliable WISE detections in
at least one band (96\% of whole SDSS DR7 quasar catalog). To get a more
reliable estimate of the bolometric luminosity and mid-infrared luminosity,
we require that the SDSS spectra cover the rest-frame wavelengths from
2000--4000~\AA\AA~ and that the WISE bands access to at least 10$\mu$m in
the quasar rest frame. This leads to a redshift cut $0.76\leq z\leq1.17$.
17639 quasars fall in this redshift range.

We fit the rest-frame SDSS spectra, corrected for the Galactic extinction
using the dust map of \citet{schlegel.etal.1998} and the extinction curve
of \citet{fitzpatrick.1999}, in three relative line-free windows
2150--2200, 3030--3100 \& 4150--4250~\AA\AA~ \citep[see][]{vanden.berk.etal.2001}
with a power-law function
\begin{equation}
	F_{\lambda} = c \, (\lambda/3000)^{\alpha} ,
\end{equation}
where normalization factor $c$ and spectral slope $\alpha$ are two 
free parameters to be determined by standard $\chi^2$ minimization 
approach\footnote{We use the MPFIT package for all nonlinear fitting 
in this work.}. Bad pixels are removed using the mask bits in the SDSS 
spectrum. To achieve reliable fitting, we further require that the 
number of good pixels must be greater than 50, and that the median
signal-to-noise ratio ($S/N$) should be greater than 3 in each window.
1220 sources (7\%) fail to meet these criteria and are thus dropped.
Another 71 sources with broad absorption lines \citep[according to][]
{shen.etal.2011} were also excluded. Furthermore, we discount another
69 sources with fitted spectral slope $\alpha$ greater than 0 because
their continua are severely reddened and are likely contaminated by 
starlight. By this stage, 16275 sources (92\%) are left.

According to \citet{shen.etal.2011}, 15279 of these sources lie in the 
footprint of the Faint Images of the Radio Sky at Twenty-Centimeters 
\citep[FIRST;][]{white.etal.1997} survey. And 1568 of them are classified 
as `radio-loud quasars'.
In the following analysis, we will examine the potential differences between
radio-loud and radio-quiet quasars on the properties we are discussing.
Furthermore, since our purpose is to study the mid-infrared emission of warm
dust, we will first restrict our analysis to the primary sample of 12483 out
of 16275 quasars that have reliable detections in ALL four WISE bands, and
then examine the properties of the rest 3792 quasars which only have upper 
limits of W3 or/and W4 magnitudes. All information for the entire sample 
relevant to this paper is available online as the supplementary material 
(see Appendix and Table~\ref{table} for a description).

\subsection{Near-UV and Mid-Infrared flux}
\label{flux}

We calculate the integrated NUV continuum flux, $\FNUV$, in the rest-frame
wavelength range of 2000--4000~\AA\AA~ by simply integrating the best fitted
power-law model. It excludes the reprocessed components such as Balmer and
pseudo Fe II continua, and broad emission lines. The errors of $\FNUV$ come
from two sources: the uncertainties of fitting parameters $c$ and $\alpha$, and
the uncertainty of SDSS spectrophotometric calibration. Using error propagation,
we estimate that the former introduces only a small error, typically $0.5\%$.
We expect the latter to be the dominant source. To quantify this, we notice
that the difference between spectrophotometric and photometric magnitudes has 
a 1$\sigma$ scatter of 0.05 mag at $r$ 
band\footnote{see http://www.sdss.org/dr7/products/spectra/spectrophotometry.html}, 
which is mostly attributed to the uncertainty in spectroscopic calibration. 
This indicates the spectral flux at $r$ band has a relative uncertainty of 
$\sim0.047$. We will quote 5\% in $\FNUV$ as the calibration uncertainty hereafter.

For mid-infrared (MIR) emission, we will focus on the integrated flux in the 
quasar's rest-frame wavelength range of 3--10~$\mu$m. This regime is covered 
by the WISE W2, W3 \& W4 bands for our chosen redshift range. Furthermore, 
the infrared emission in this regime is primarily dominated by the warm and 
hot dust heated by quasars. Finally, the mean quasar SED in this wavelength 
range is relatively smooth according to \citet{richards.etal.2006} and 
\citet{shang.etal.2011}. Therefore, it is relatively straightforward to 
obtain the integrated flux from the WISE photometric data without invoking 
detailed dust modelling.

In practice, we approximate the MIR SED as a broken power-law with broken 
wavelength at the effective wavelength of W3 band at the rest frame of the 
quasar. Then we integrate the model from rest-frame 3~$\mu$m to 10~$\mu$m, 
to obtain the integrated MIR flux $\FMIR$. To estimate the uncertainty of 
this measured MIR flux, we apply the same approach to every object in the 
sample of \citet{shang.etal.2011}, which have the \textit{Spitzer} MIR spectra. 
We find that our approach gives $\FMIR$ 3\% higher on average than the 
direct integration of IR spectra over the correspondent wavelength range, 
with a scatter of 8\%. Hereafter, we adopt 3\% as the global offset and 
8\% as the uncertainty introduced by our method. Finally, we combine this 
uncertainty with statistical uncertainty comes from the uncertainty of 
WISE magnitudes as the final uncertainty of $\FMIR$. The spectra slope 
$\alpha$, NUV flux $\FNUV$ and MIR flux $\FMIR$ for the entire sample can 
be found in the online table (see Table~\ref{table} for a reference).

\section{Extinction Correction}
\label{extinction}

\subsection{Correlation between $\FMIR/\FNUV$ and Spectral Slope $\alpha$}
\label{correlation}

\begin{figure}
  \includegraphics[width=85mm]{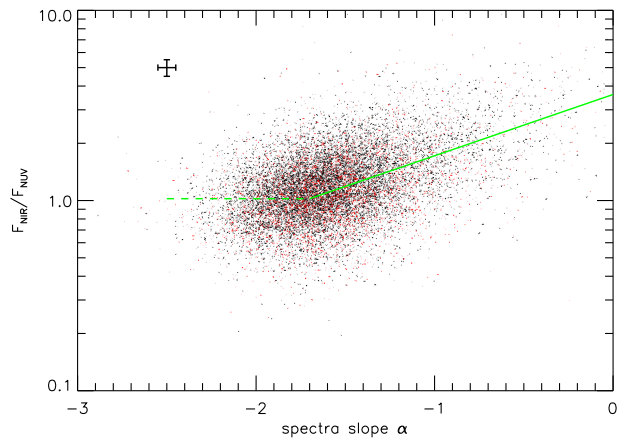}
  \caption{Correlation between $\FMIR/\FNUV$ and spectra slope $\alpha$.
	The black dots represent the 12483 objects in our primary sample 
	and red dots represent the 3792 objects with only upper limits of 
	W3 or/and W4 magnitudes. The green line displays the tendency 
	predicted by SMC-like extinction, only for those $\alpha>-1.7$. 
	The green dash line denotes the mean value of $\FMIR/\FNUV$ for 
	quasars with $\alpha<-1.7$.}
  \label{fig1}
\end{figure}

The observed near-ultraviolet spectral slopes of quasars in our sample lie in a
quite wide range from $-2.9$ for the bluest one to $0$ for the reddest one, with a
median value of $-1.7$. 
We plot $\FMIR/\FNUV$ versas $\alpha$ in Fig.~\ref{fig1}. There is a fairly strong 
correlation between $\FMIR/\FNUV$ and $\alpha$ with a Spearman ranking correlation 
coefficient $R_s=0.41$, corresponding to a chance probability of $P_r<10^{-6}$.
We fit this relation with a linear function\footnote{We use the IDL procedure
robust\_linefit.pro for all the linear fitting in this work.}
\begin{equation}
	\log(\FMIR/\FNUV)=k\cdot\alpha + b ,
\end{equation}
where $k$ and $b$ are free parameters. For all 12483 quasars including in our 
primary sample, the best fit yields $k=0.24\pm0.02$ and $b=0.45\pm0.05$.

This correlation can be naturally interpreted as wavelength-dependent dust
extinction. For example, \citet{richards.etal.2001} showed that SDSS colors 
of quasars can be well represented by power laws of spectral indexes 
$\alpha=-1.5\pm0.65$ with a red tail possibly due to dust extinctions. 
Because extinction is larger at shorter wavelengths, it makes the observed
optical-UV spectrum flatter (redder), and simultaneously raises observed
$\FMIR/\FNUV$. The large intrinsic scatters of $\alpha$ and $\FMIR/\FNUV$
dominate the variations at small $\alpha$ and small $\FMIR/\FNUV$, thus
make the correlation weaker for blue sources as seen in Figure~\ref{fig1}.
Given that most blue quasars are likely not dust reddened, we check how the
parameter $k$ changes with different $\alpha$ ranges. For the 7700 sources 
with $\alpha>-1.7$, we obtain $k_{1.7}=0.29\pm0.02$, and for the 4505 sources 
with $\alpha>-1.5$, the result is $k_{1.5}=0.33\pm0.02$.

By comparing the behaviours of radio-loud and radio-quiet quasars in our 
sample, we find that this $\FMIR/\FNUV$--$\alpha$ relation is independent 
of radio activities. Also, for the 3792 sources that are removed from our 
primary sample, we calculate their $\FMIR$ using the same approach described 
in \S~\ref{flux} with the upper limits of W3 \& W4 magnitudes provided in the 
WISE catalog, and find that they share the same behaviour as for the primary 
sample (the red dots in Fig.~\ref{fig1}). Therefore, we will not distinguish 
these different categories of quasars when analyzing extinction effect below.

Previous studies suggested that dust extinction to quasars can be 
characterized by an SMC-like extinction curve
\citep[e.g.][]{richards.etal.2003,hopkins.etal.2004,gaskell.etal.2004}.
To illustrate the extinction effect on $\FMIR/\FNUV$--$\alpha$ relation
quantitatively, we apply SMC-like extinction \citep{pei.1992} of different 
values of $E(B-V)$ to an intrinsic blue QSO continuum, which is described 
by an intrinsic spectral slope $\beta$ ($F_\lambda\propto\lambda^\beta$) 
in the NUV range and a certain fraction of reprocessed infrared emission, 
to get a set of faked quasar continuua. We measure the NUV spectral slope
($\alpha$) and $\FNUV$ for each faked continuum in exact the same way as 
that described above for real QSOs. It turns out that the relationship
between $\alpha$ and $E(B-V)$ can be well described by a linear function,
\begin{equation}
	E(B-V)=c\,(\alpha-\beta), \label{ebv}
\end{equation}
for $E(B-V)<0.3$. Particularly, for $\beta=-2.0$, we have $c=0.127$.
For such a $E(B-V)$ range, the dust extinction has little effect on the
$\FMIR$, which can thus be considered as a constant. The relation between
$\log(\FMIR/\FNUV)$ and $\alpha$ for the mock quasars can be well described 
by a linear function with a slope of 0.322, independent of the value of 
$\beta$ for a wide range of $-2.2<\beta<-1.8$. This slope is also very 
close to the observed one $k_{1.7}=0.29$ or $k_{1.5}=0.33$. Therefore, 
our analysis supports that the observed $\FMIR/\FNUV$--$\alpha$ relation
is statistically attributed to the dust extinction. And the scattering of 
this correlation is dominated by intrinsic dispersions of the continuum
slope and $\FMIR/\FNUV$ (green lines in Fig.~\ref{fig1}).

\subsection{Extinction Correction for $\FNUV$}

In this subsection, we will correct $\FNUV$ for dust extinction in
a statistical way using the relation derived in the last section.
First, we estimate the intrinsic distribution of $\log(\FMIR/\FNUV)$
using blue quasars with $\alpha<-1.7$. As mentioned above, in this regime,
extinction is small, and variations of $\log(\FMIR/\FNUV)$ and $\alpha$
are dominated by intrinsic scatters, rather than reddening effect, albeit
the latter must be present. The distribution of $\log(\FMIR/\FNUV)$ for
blue quasars can be well described by a Gaussian function with a mean
value of 0.010 ($\FMIR/\FNUV=1.02$), and a dispersion of $\sigma=0.144$.

As we have shown in the last subsection, the dependence of $\FMIR/\FNUV$
on observed spectral slope $\alpha$ can be explained by dust extinction.
Therefore, it is possible to correct $\FNUV$ in a statistical way for
the extinction using the derived $\FMIR/\FNUV$--$\alpha$ relation, 
with fitted spectral slope $\alpha$ as an extinction indicator.
Assuming the mean $\FMIR/\FNUV$ at different $\alpha$ is the same (1.02) 
as that for blue quasars, and following the $\FMIR/\FNUV$--$\alpha$ relation
derived in the last subsection, we calculate the correction factor for $\FNUV$:
\begin{equation}
	C= \left\{
	  \begin{array}{ll}
	    3.5 \times 10^{0.322 \times \alpha} & \mbox{if $\alpha>-1.7$} \\
	    1  & \mbox{if $\alpha<-1.7$} .
	  \end{array}
	  \right.
\end{equation}
In the following analysis, we still use the notation $\FNUV$ to denote
the NUV flux, while it has already been corrected for extinction effect.
It should be pointed out that this correction is only in statistical sense 
for the entire sample rather than for individual object since the intrinsic
UV continuum slope $\beta$ and $\FMIR/\FNUV$ have fairly broad distributions.
Nevertheless, by applying different cut on $\alpha$ as the standard for blue
quasars, we estimate the uncertainty of this correction for individual object
is less than 10\%.

\section{Warm Dust Covering Factor and its Relation with AGN Properties}
\label{relation}

\subsection{Covering Factor of Warm Dust}
\label{wd}

The observed MIR flux $\FMIR$ comes from three components: emission of the
warm and hot dust heated by AGN, young-star emission at MIR band, and the
emission of the accretion disc extending to MIR bands. Star-formation galaxies 
are characterized by strong PAH emission in MIR. Previous infrared spectroscopic
studies suggest that PAH features are very weak in quasar spectra. Therefore,
this component is not statistically important in our MIR range of rest-frame
3--10~$\mu$m. We will neglect it in the analysis hereafter. We estimate the
accretion disc component by extrapolation of the optical/UV power-law corrected
for reddening effect. This could be considered as an upper limit considering
spectral flattening in the near-infrared due to starlight contribution in the
SDSS spectrum\footnote{Starlight contribution in our continuum windows is general 
small for luminous quasars in this paper \citep[e.g.][]{stern.laor.2012}.}.
Numerically, the disc contribution to $\FMIR$ is 5.2\% of $\FNUV$ for
$\beta=-2.2$, 9.3\% of $\FNUV$ for $\beta=-2.0$ and 16.6\% of $\FNUV$
for $\beta=-1.8$. Since we focus on the MIR emission of the dust tori, 
we will subtract 0.093$\FNUV$ from $\FMIR$ to remove the contribution of the 
accretion disc and the slight starlight. The remained value, also denoting 
by $\FMIR$, should be considered as the MIR emission of the warm dust.

Assuming that the NUV and MIR are isotropic, we can calculate the NUV
luminosity $\LNUV$ and MIR luminosity $\LMIR$ from $\FNUV$ and $\FMIR$,
respectively. Next, we consider the bolometric correction ($BC$) for the 
total luminosity from optical to X-ray. Since we have already corrected 
for extinction effect, we will use the mean SED of optically blue quasars
in \citet{richards.etal.2006} to calculate this correction. We calculate 
the $\LNUV$ using the same method described in \S~\ref{flux}, and integrate
$\log(\nu f_\nu)$ over $\log\nu$ from 1~$\mu$m to 10~keV to obtain the total
luminosity in this range\footnote{We discount the IR luminosity here since
most mid- and far-infrared emission is the re-radiation of the absorbed
continuum \citep{shang.etal.2011}. But this still includes a small contribution
from emission lines, which is doubly counted.}.
This yields $BC=4.34$. Thus bolometric luminosities can be calculated by
$\Lbol=BC\times\LNUV$. Finally, we define the covering factor of warm dust
as $\CF=\LMIR/\Lbol$. Our definition of covering factor is similar to that 
of \citet{calderone.etal.2012}, but different from theirs in the sense that 
we consider the IR flux in a fixed MIR range dominated by warm dust rather 
than the whole IR bands covered by WISE. Besides, we do not take into account 
anisotropic emission intentionally by this stage, while we will examine 
orientation effect in the last section.

\subsection{Dependence of $\CF$ on AGN Properties}

\begin{figure*}
  \includegraphics[width=17cm]{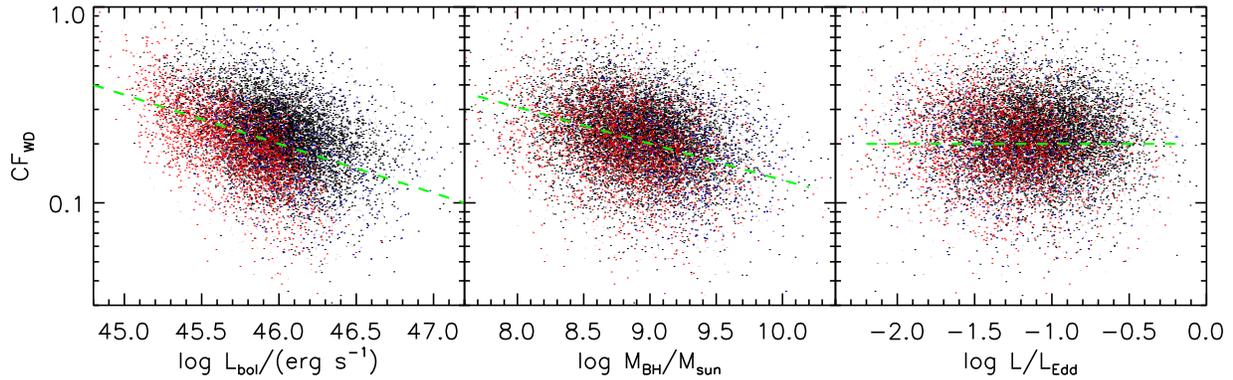}
  \caption{Relations between warm dust covering factor $\CF$ and AGN properties:
	bolometric luminosity $\Lbol$ (left panel), black hole mass $\MBH$ 
	(middle panel), and accretion rate $\Lbol/L_{\rm Edd}$ (right panel). 
	The meanings of black and red dots are exactly the same as in figure 1. 
	The blue dots represent radio-loud quasars in the primary sample. 
	The green dash lines represent the linear fit for these relations.}
  \label{fig2}
\end{figure*}

In this subsection, we consider the relations between $\CF$ and AGN
properties: bolometric luminosity $\Lbol$, black hole mass $\MBH$,
and Eddington ratio $L/L_{\rm Edd}$. We present these relations for 
our sample in Fig.~\ref{fig2}. We adopt the virial black hole masses 
in the catalog of \citet{shen.etal.2011} with an additional extinction 
correction. These masses are estimated using the width of 
MgII $\lambda 2798$ and UV continuum luminosity at 2800~\AA~ 
\citep[Eqn. (2) in][]{shen.etal.2011}.
\begin{eqnarray}
	\log(\frac{M_{\rm BH,vir}}{M_{\odot}}) & = & a +
		b\,\log\,(\frac{\lambda L_{\lambda}}{10^{44}\, \rm erg\,s^{-1}}) \nonumber \\
	& &	+2\,\log(\frac{\rm FWHM}{\rm km\,s^{-1}}) ,
\end{eqnarray}
with $a=0.740$ and $b=0.62$. \citet{shen.etal.2011} did not consider 
intrinsic extinction on the UV luminosity. In the following analysis, 
we will apply a correction by adding
\begin{equation}
	\Delta=1.588*E(B-V)
\end{equation}
to the value provided by \citet{shen.etal.2011}. This correction is calculated 
according to SMC reddening curve at 2800~\AA. And $E(B-V)$ can be calculated 
from spectra slope ($\alpha$) using Eqn.~(\ref{ebv}) in \S~\ref{correlation}. 
Bolometric luminosities and black hole masses for the entire sample are also 
provided in the online table.

First, $\CF$ is anti-correlated with $\Lbol$. The Spearman rank 
correlation coefficient ($R_s$) of this correlation is $-0.36$, 
indicating the anti-correlation is very significant ($P_r<10^{-6}$). 
A linear fit yields
\begin{eqnarray}
	\log\CF & = & (-0.23\pm0.02)\log\Lbol/(\rm erg\cdot s^{-1}) \nonumber \\
	& &	+ (9.79\pm0.04) .
\end{eqnarray}

There is a significant anti-correlation between $\CF$ and black 
hole mass $\MBH$ with a Spearman rank correlation coefficient
($R_s=-0.27$, $P_r<10^{-6}$). A linear fit yields
\begin{eqnarray}
	\log\CF & = & (-0.13\pm0.02)\log(M_{\rm BH}/M_{\odot}) \nonumber \\
	& & + (0.50\pm0.02) .
\end{eqnarray}
Finally, no apparent correlation is found between $\CF$ and Eddington ratio,
\begin{equation}
	\frac{L}{L_{\rm Edd}}=\frac{L_{\rm bol}}{1.5\times10^{38}M_{\rm BH}} 
\end{equation}
($\rho=-0.008$ and $P_r=0.34$).

We compare radio-quiet and radio-loud quasars on these three diagrams,
and find that they are indistinguishable. Particularly, we have fitted
related parametres separately for the radio-loud subsample. First, the
median value of $\CF$ is consistent with that of the radio-quiet subsample
within 10\%. Second, the slopes of $\CF$--$\Lbol$ and $\CF$--$\MBH$ relations 
are $-0.21\pm0.02$ and $-0.15\pm0.02$, respectively. The distribution of $\CF$,
and these fitting slopes are consistent with their 1 $\sigma$ uncertainties.
Therefore, we conclude that none of these three relations are affected by
radio activities.

Next, we consider the 3792 sources that only have upper limits of W3 and/or 
W4 magnitudes. From the left panel of Fig~\ref{fig2}, it is obvious that these 
objects have systematically lower $\Lbol$, while at the same time their $\CF$ 
(upper limits) are similar to those of quasars in our primary sample, so they 
shift to the left on the $\CF$--$\Lbol$ diagram. In other words, they show
systematically lower $\CF$, comparing to the quasars of similar luminosities 
in the primary sample. Because the fraction of these sources is small, we get  
almost an identical slope for $\CF$--$\Lbol$ relation for the entire sample. 
From the middle and right panels of Fig.~\ref{fig2}, these quasars have similar 
distribution of black hole masses, but systematically lower $L/L_{\rm Edd}$ 
than the primary sample. However, they follow the same $\CF$--$\MBH$ and 
$\CF$--$L/L_{\rm Edd}$ relations as the primary sample.

\section{Discussion and Conclusion}
\label{discussion}

\subsection{On Radio-loud Quasars}

Approximately 10\% objects in our sample are radio-loud quasars. As we have
discussed earlier, they exhibit the same correlations as radio-quiet quasars.
Therefore, the majority of the radio-loud quasars have the very similar structure 
of accretion disc and dusty torus. This is consistent with \citet{richards.etal.2006} 
and \citet{shang.etal.2011} that mean SEDs of radio-quiet and radio-loud quasars 
in NUV and MIR are quite similar.
However, it is of necessity to point out that a small fraction of the radio-loud
quasars within our sample are blazars, in which the relativistic jets are beamed
toward us. In this case, the relativistic jets would contribute significantly or
even dominate the observed flux at near-ultraviolet and mid-infrared bands.
Therefore, NUV and MIR emission in these objects will no longer represent the 
primary emission of the accretion discs and re-emission of the dusty tori.
Thus our definition of warm dust covering factor is no long appropriate for
these blazars. Nevertheless, the fraction of blazars in our whole sample is
quite small ($\sim1\%$ of type 1 quasars according to the argument of 
\citealt{padovani.2007}), they will definitely not affect our conclusion.

\subsection{Comparison with Previous Studies}

In this work, we use the mid-infrared to near-UV luminosity ratio to estimate
the covering factor of warm dust, and find a decrease of $\CF$ with increasing
$\Lbol$. Our results are in qualitative consistency with previous works.

\citet{maiolino.etal.2007} estimated the dust covering factor for a few tens of 
QSOs using monochromatic flux ratio of 6.7$\mu$m to 5100\AA, which is similar to 
our method in the spirit. However, their sample was constructed by merging two 
groups of QSOs from totally separate redshift ranges. Moreover, their analogous 
CF--$\Lbol$ correlation is far less significant when only considering either one 
group of objects with close redshifts. Therefore, their result might be hampered 
by possible evolution effect. 
A similar trend was obtained by \citet{treister.etal.2008}, who made use of 
nearly one hundred objects constructed from different surveys. In comparison 
with these works, our sample is much larger, and is uniformly selected in a 
relatively small redshift range.

More recently, \citet{mor.trakhtenbrot.2011} used the preliminary data release of
WISE to study the hot dust component for a large and uniformly selected sample.
Their definition of hot dust covering factor is quite similar to ours, and their
results also revealed a dependence of hot dust CF on bolometric luminosity.
Their sample covers a very wide redshift range and their hot dust luminosities 
were estimated using a certain dust model. Instead, we take a model-independent 
approach in our work.
Similarly, \citet{calderone.etal.2012} studied the dust covering factor for all
radio-quiet AGNs constructed from WISE and SDSS within a small redshift range.
These authors reported a similar trend of CF--$\Lbol$ relation by dividing the
whole sample into three bins. Comparing with these studies, our work has taken
the extinction effect into account. Also, we used the extinction-corrected broad
band NUV luminosity, which is a better estimation of bolometric luminosity.
Furthermore, we find that CF is significantly correlated with black hole mass,
but is independent of Eddington ratio. The latter results are consistent with
those reached by \citet{cao.2005} for PG quasars.

In the literature, many authors used the fraction of type 2 objects as an
indicator of total dust covering factor and reached a similar conclusion.
For instance, \citet{simpson.2005} and \citet{hao.strauss.2004} found a 
similar dependence of type 2 fraction on the [O III] luminosity using a 
magnitude-limited sample from the SDSS spectroscopic survey. 
Similar results were reported by other authors using sample constructed 
from flux-limited X-ray survey \citep[e.g.][]{hasinger.2008} or a combination 
of multi-band survey \citep{hatziminaoglou.etal.2009}. However, some other 
authors have reported the opposite results. \citet{lawrence.elvis.2010} 
suggested that there is no correlation between type 2 fraction and [O III] 
luminosity after removing LINERs from the definition of type 2 objects, 
and so did \citet{lu.etal.2010} by taking into account of various selection 
effects. A caution should be taken when comparing our results with those 
derived from the fraction of type 2 AGNs. We only consider the covering 
factor of warm dust, which is merely a portion of the obscuring material. 
Furthermore, a fraction of the type 2 objects are obscured by dust in their host 
galaxies rather than dust tori themselves \citep{goulding.alexander.2009}.

\subsection{Effects of Anisotropic Continuum}

It should be emphasized that up to now, our whole analysis is based on the
assumption that the NUV and MIR emission of quasars are isotropic. However,
the NUV emission from a geometrically thin but optically thick accretion disc
is predicted to be stronger at polar directions than at equatorial directions
\citep[e.g.][]{shakura.sunyaev.1973,laor.netzer.1989,fukue.akizuki.2006}.
If this is the case, our estimation of bolometric luminosity would be biased 
depending on the viewing angle \citep[cf.][]{nemmen.brotherton.2010,runnoe.etal.2012}
and our definition of covering factor would certainly be inaccurate. For example,
if an object is viewing relatively edge-on, we will underestimate its bolometric
luminosity $\Lbol$ while overestimate its covering factor $\CF$. In a statistical
view, this can result in an anti-correlation between $\CF$ and $\Lbol$, the same
tendency as we have discovered from our sample.

To illustrate this more clearly, we construct a simple model to simulate the
effect of different viewing angles. The simulation involves the same number of
faked quasars as for our sample. We adopt three assumptions: (1) The intrinsic
$\CF$ and $\Lbol$ are independent, and obey log-normal distributions respectively.
(2) The inclinations are random in a range of $1/2<\mu<1$, where $\mu=\cos\theta$
(the viewing angle). A lower limit on $\mu$ is set, since we require the
line-of-sight must lie in the open angle of the torus, otherwise the NUV emission
would be totally absorbed. The value 1/2 is corresponding to a type 2 to type 1
ratio of 1:1\footnote{We set the lowest $\mu$ to the value for the average torus
oppening angle, rather than that giving by the warm dust covering factor of each
object because the covering factor of thick cold and hot dust is not known.}
\citep{lawrence.elvis.2010,reyes.etal.2008}. (3) The observed $\Lbol$ and $\CF$ 
for each object are calculated from observed NUV flux and NIR/NUV flux ratio 
viewing from a certain direction. The dependence of NUV flux on viewing angle 
used in our model is
\begin{equation}
	f(\mu)=f_0\times\mu\,(\mu+\frac{2}{3}) ,
\end{equation}
following the results of \citet{fukue.akizuki.2006} who have taken into account 
limb darkening effect. Moreover, we require that the observed $\Lbol$ and $\CF$ 
have exactly the same distribution as for our sample.

\begin{figure}
  \includegraphics[width=85mm]{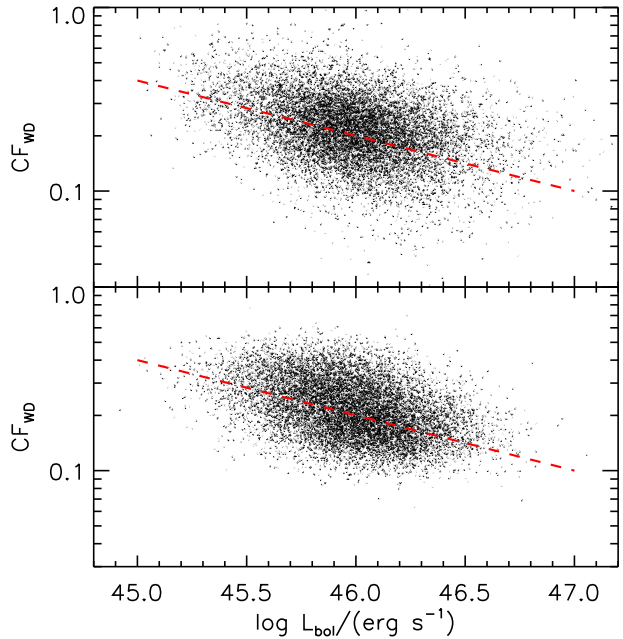}
  \caption{Above: the $\CF$--$\Lbol$ relation for the primary sample.
	Bottom: the same relation for the simulated sample.}
  \label{fig3}
\end{figure}

We plot our simulated $\CF$--$\Lbol$ relation in Fig.~\ref{fig3}, in comparison 
with the same relation of the sample. It turns out that our simulation obtains 
a similar but more compact relation. This seems to indicate that the observed 
$\CF$--$\Lbol$ may result from anisotropic NUV continuum.

However, we stress that our simulation is initial and needs to be refined 
in several aspects. First, the assumption that the viewing angles are
uniformly distributed is not satisfied by a flux-limited sample, such as 
the SDSS quasar sample used here. For example, a quasar close to the flux 
limit may move out of this limit if it is observed at a higher inclination.
Thus it will be missed in the sample. In other words, the sample is biased
against least luminous quasars at large inclinations.
We note that the continuum flux decreases by a factor of 2.9 from $\mu=1$ to
$\mu=1/2$, so this selection effect will significantly affect the inclination
distribution of quasars with luminosities of a factor of 2.9 above the minimum
value. The net effect will weaken the $\CF$--$\Lbol$ correlation.
In principal, we can incorporate this selection effect into the simulation
by using the quasar luminosity function. However, we should understand that
dust extinction is also coupled with the selection effect in the real case.
Furthermore, dust extinction may or may not correlate with the inclination.
If the absorber is the continuous extension of the dusty torus, there will
certainly be a correlation; while on the other hand, if the absorption is due
to the interstellar medium of the host galaxy, there may be no correlation.
Because of lack of such detailed knowledge, we did not carry more complicated
simulations.

Second, IR emission is also likely to be anisotropic due to the radiation
transfer effect. According to the calculations of 
\citet{nenkova.etal.2008a,nenkova.etal.2008b},
for a typical clumpy dusty torus model, the MIR flux is $\sim50\%$ weaker
viewing from a 45 degree inclination than viewing from face-on direction.
This factor is very similar to the anisotropy of NUV flux. In that case,
the $\CF$--$\Lbol$ relation will be much shallower due to the decrease of
infrared flux viewing from higher inclinations. Hence the correlation will
be weakened.

Therefore, although our simulation suggests that the observed $\CF$--$\Lbol$
relation may be attributed to the inclination effect, there is fairly possible
that the relation is intrinsic. A detailed study combining anisotropic radiation
from both the discs and dust tori is very favourable in the future.
Also, continuum variation on time scales of years will cause similar trend due
to a delayed response of dust emission. The fact that CF is correlated with
black hole mass but not correlated with Eddington ratio further indicates that
luminosity dependence of CF is not solely due to the anisotropic emission alone.

To summarize, we use a complete sample in a narrow redshift range to study the warm
dust emission of $\sim16,000$ quasars. We correct for the dust extinction effect in
a statistical way for the first time in this kind of study. We find that $\CF$ is
correlated with $\Lbol$ and $M_{BH}$, while not with Eddington ratio. Monte Carlo
simulation suggests that $\CF$--$\Lbol$ relation may be significantly affected by
anisotropic disc emission. But it may not be the dominant factor for this relation.

\section*{Acknowledgments}

We thank the anonymous referee for the comments regarding radio-loud quasars
and potential limitations of the assumptions using in the simulation,
which largely helped us improve our work. This work was supported by
the Chinese NSF through NSFC-10973013 and NSFC-11233002.
This work has made use of the data obtained by SDSS and WISE.
Funding for the SDSS and SDSS-II has been provided by the
Alfred P. Sloan Foundation, the Participating Institutions,
the National Science Foundation, the U.S. Department of Energy,
the National Aeronautics and Space Administration, the Japanese Monbukagakusho,
the Max Planck Society, and the Higher Education Funding Council for England.
The SDSS web site is http://www.sdss.org/.
WISE is a joint project of the University of California, Los Angeles,
and the Jet Propulsion Laboratory of California Institute of Technology,
funded by the National Aeronautics and Space Administration.
The WISE web site is http://wise.astro.ucla.edu/.

\section*{Appendix}

We present the primary data of all 16275 sources in our sample with an ASCII table 
as the supplementary material of this paper. The entire table is available online. 
Here we offer a small portion of the table for a guidance (Table~\ref{table}).

\begin{table*}
\begin{center}
\caption{Primary data of our sample.} \label{table}

\begin{tabular}{cccccccccc}

\hline\hline
Name   & Redshift & Spectra slope & NUV flux & MIR flux & $\CF$ &
        Bolometric luminosity & Black hole mass & Flag$^b$ & FIRST \\
(SDSS J) & ($z$) & ($-\alpha$) & ($\FNUV$$^a$) & ($\FMIR$$^a$) &  &
        $\log(\Lbol/(\rm erg\,s^{-1}))$ & $\log(\MBH/M_{\odot})$ &  & flag$^c$ \\

\hline
000013.14+141034.6 & 0.958 & 1.657 & 2.773e-13 & 2.617e-13 & 0.217 & 45.757 & 8.800 & 0 & -1 \\
000024.02+152005.4 & 0.989 & 1.744 & 2.820e-13 & 2.044e-13 & 0.167 & 45.798 & 9.181 & 1 & -1 \\
000025.21+291609.2 & 0.924 & 1.718 & 8.433e-14 & 9.096e-14 & 0.249 & 45.201 & 8.216 & 0 & -1 \\
000025.93+242417.5 & 1.156 & 1.296 & 1.487e-13 & 2.077e-13 & 0.322 & 45.688 & 9.310 & 1 & -1 \\
000026.29+134604.6 & 0.768 & 1.691 & 3.206e-13 & 1.291e-13 & 0.093 & 45.582 & 8.670 & 1 & -1 \\
000028.82-102755.7 & 1.152 & 1.277 & 4.372e-13 & 2.150e-13 & 0.113 & 46.152 & 9.449 & 0 &  1 \\
000031.86+010305.2 & 1.093 & 1.086 & 2.274e-13 & 3.818e-13 & 0.387 & 45.812 & 8.552 & 0 &  0 \\
000042.89+005539.5 & 0.951 & 1.805 & 9.408e-13 & 1.033e-12 & 0.253 & 46.279 & 8.844 & 0 &  0 \\
000057.67-085617.0 & 1.095 & 1.460 & 8.727e-13 & 8.643e-13 & 0.228 & 46.397 & 8.778 & 0 &  0 \\
000123.96-102458.1 & 1.058 & 1.616 & 6.734e-13 & 6.254e-13 & 0.214 & 46.248 & 9.338 & 0 &  0 \\
000123.98+284249.4 & 0.936 & 0.269 & 2.419e-13 & 2.935e-13 & 0.280 & 45.672 & 9.255 & 0 & -1 \\
000140.70+260425.5 & 0.765 & 1.385 & 3.656e-13 & 3.700e-13 & 0.233 & 45.636 & 8.670 & 0 & -1 \\
\hline\hline
\multicolumn{10}{l}{The full table including all 16275 sources is available online as supplementary material.
        A portion is shown here for guidance of its content.} \\
\multicolumn{10}{l}{$^a$ The flux is in unit of $\rm erg\,s^{-1}\,cm^{-2}\,\AA^{-1}$.} \\
\multicolumn{10}{l}{
	Here $\FNUV$ refers to the value after extinction correction (see \S~\ref{extinction}) and
        $\FMIR$ refers to the value after discounting the disc contribution (see \S~\ref{wd}).} \\
\multicolumn{10}{l}{$^b$ Flag: 0=12483 source in our primary sample;
        1=the rest 3792 sources which only have upper limits for W3 or W4 magnitudes.} \\
\multicolumn{10}{l}{$^c$ First flag: the same as column 15 in the catalog of \citet{shen.etal.2011}.
         -1=not in FIRST print; 0=FIRST undetected; 1=core dominant; 2=lobe dominant.} \\

\hline\hline
\end{tabular}

\end{center}
\end{table*}

\bibliographystyle{mn2e}

\label{lastpage}

\end{document}